\documentclass[12pt,fleqn]{article}

\usepackage{setspace}
\usepackage{float}
\usepackage{fullpage}
\usepackage{float}
\usepackage{wrapfig}
\usepackage{amsthm,dcolumn,graphicx,multicol,fancyhdr}
\usepackage{latexsym,subfig,graphicx,ifthen}
\usepackage{rotating,amsfonts, color,amssymb,amsmath,amscd,array,enumerate,afterpage}
\usepackage{hyperref}
\usepackage{bm}
\usepackage{titling}
\usepackage[superscript,biblabel]{cite}

\newcolumntype{P}[1]{>{\centering\arraybackslash}p{#1}}

\def\keywords{\vspace{.5em}
	{\textbf{Keywords}:\,\relax%
	}}

\title{A Robust Interrupted Time Series Model for Analyzing Complex 
Healthcare Intervention Data}
\author{Maricela Cruz\thanks{Department of Statistics, University of California, Irvine, USA}, Miriam Bender \thanks{Nursing Science Program, University of California, Irvine, USA}, Hernando Ombao \thanks{Department of Statistics, University of California, Irvine, USA} \thanks{Statistics Program, King Abdullah University of Science and Technology (KAUST), Saudi Arabia} }
\date{\vspace{-9ex}}

\begin{document}

\maketitle
\doublespacing
\begin{abstract}

%% 248 words, limit is 250 for Stats in Medicine
Current health policy calls for greater use of evidence based care delivery services to improve patient quality and safety outcomes.  Care delivery is complex, with interacting and interdependent components that challenge traditional statistical analytic techniques, in particular when modeling a time series of outcomes data that might be ``interrupted" by a change in a particular method of health care delivery. Interrupted time series (ITS) is a robust quasi-experimental design with the ability to infer the effectiveness of an intervention that accounts for data dependency. Current standardized methods for analyzing ITS data do not model changes in variation and correlation following the intervention. This is a key limitation since it is plausible for data variability and dependency to change because of the intervention. Moreover, present methodology either assumes a pre-specified interruption time point with an instantaneous effect or removes data for which the effect of intervention is not fully realized. In this paper, we describe and develop a novel `Robust-ITS' model that overcomes these omissions and limitations. The Robust-ITS model formally performs inference on: (a) identifying the change point; (b) differences in pre- and post-intervention correlation; (c) differences in the outcome variance pre- and post-intervention; and (d) differences in the mean pre- and post-intervention.  We illustrate the proposed method by analyzing patient satisfaction data from a hospital that implemented and evaluated a new nursing care delivery model as the intervention of interest. The Robust-ITS model is implemented in a R Shiny toolbox which is freely available to the community.

\end{abstract}


\section{Introduction}
Current health policy calls for greater use of evidence-based practice (EBP) in delivering healthcare services to improve patient outcomes \cite{InstituteofMedicineUSRoundtableonEvidenceBasedMedicine:2009fu}. 
In this paper, we develop a robust time series model for estimating the impact of an intervention on health outcomes. 
The complexity of healthcare is becoming increasingly recognized: patients, providers, resources and contexts of care interact in dynamic ways to produce health outcomes that many times do not align with expectations \cite{Hawe:2015ca}. This complexity and interdependency makes it difficult to assess the true impact of interventions designed to improve patient healthcare outcomes in terms of research design and statistical analysis \cite{Datta:2013gf}. 
 Methodologies capable of managing data interdependency are being developed, yet are still considered less robust than traditional methods which assume that intervention causal factors can be analyzed without consideration of their participant samples and contexts \cite{Clark:2013hs}. 
 Interrupted time series (ITS) design has emerged as a quasi-experimental methodology with the strongest power to infer causality without stripping contextual and temporal factors from the analysis \cite{Shadish:2002uv}. 

Segmented regression is the most popular statistical method for analyzing time series data of healthcare interventions \cite{Taljaard:2014kc, Penfold:2013bc}. 
While powerful, there are limitations to this approach; namely that it restricts the interruption to a predetermined time point in the series or removes the set of time points for which the intervention effects may not be realized, and neglects the plausible differences in autocorrelation and variability present in the data.

In this paper, we develop the Robust-ITS model which is a novel model for interrupted time series. One advantage of the Robust-ITS model compared to previous methods is its ability to estimate, rather than assume a priori, the time when the effect of intervention initiates (the change point).
In practice, the change point may occur either \textit{before} or \textit{after}
the official intervention time. For instance, an intervention intended to improve care quality requiring a training over several months or weeks 
may already produce a change in the outcome even before the formal intervention time (before the official start of intervention) if the trainees execute their training as they learn.  

We propose a method which regards the change point as variable, appropriate for situations where the data warrants such treatment. Robust-ITS allows us to test when the effect of the intervention initiates in situations when pin pointing the change is of interest. 
Nonetheless, if the aim is to make causal inference it may be better to pre-specify the change point or remove the set of possible change points (or the set of points for which the intervention has not fully been realized) from the analysis, as in traditional segmented regression for ITS designs.

The main contributions of Robust-ITS are the formal tests for differences in the correlation structure and variability between the pre- and post-change point.

\begin{figure}[h!] \centering
	\includegraphics[width=17.5cm,height=11cm]{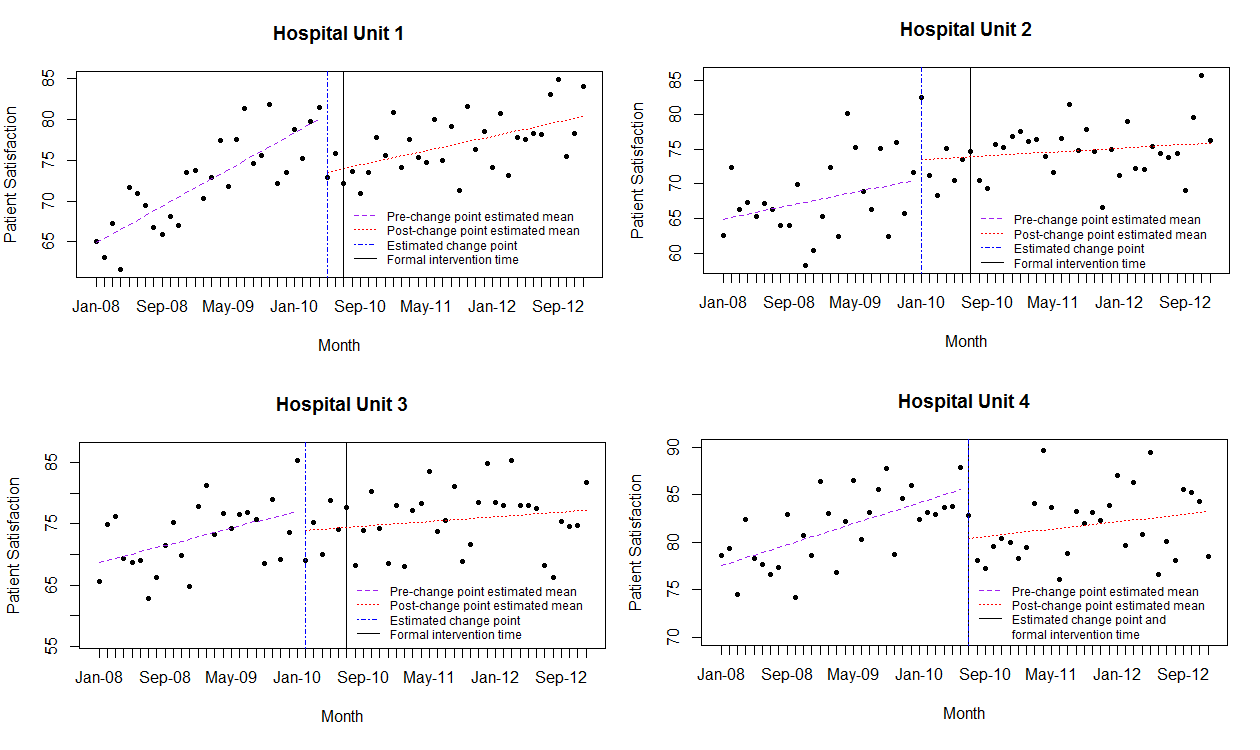} 
	\caption{Plots the time series of observed average patient satisfaction for each unit, the estimated change point, estimated means, and formal intervention time. The estimated means and change point are obtained from modeling the time series with Robust-ITS.}
	\label{UnitTS}
\end{figure}

The data used for the model development come from a study aimed to determine the influence of redesigning a nursing care delivery system on nationally endorsed quality and safety metrics \cite{ClinicalNurseLeade:2015tr}. Many nationally endorsed metrics must be publically reported, and tracked on a monthly basis via aggregate rates, counts, or ratios. The specific data used for this modeling procedure were patient satisfaction survey scores. Patient satisfaction is an important health outcome, providing a valid measure of quality of care received, and has previously been used for ITS analysis of nursing care delivery interventions \cite{Bender:2012ch}. It is also a metric that is currently being used to calculate health systems’ reimbursement for care services, via the Center for Medicaid and Medicare Services (CMS) Value Based Purchasing Program, making it a significant focus for improvement \cite{Kavaanagh:2012dt}. 
A time series plot of patient satisfaction scores from January 2008 to December 2012 at a number of units in a health care system is given in Figure \ref{UnitTS}.

The unit of analysis in the study is the care delivery microsystem, or hospital ``unit." Patient satisfaction scores are reported as aggregate scores per month, per unit. Patient satisfaction indicators include `nurse communication', `skill of the nurse', and `pain management'. Patients respond to items by selecting one of four responses: \textit{never}, \textit{sometimes}, \textit{usually}, or \textit{always}. 

For modeling we chose one outcome, the average patient satisfaction over seven patient satisfaction indicators, 
for 4 hospital units of the study setting. We refer to the average patient satisfaction score simply as patient satisfaction throughout the remainder of the paper. The intervention program, titled Clinical Nurse Leader (CNL) integrated care delivery, was the introduction of novel nursing care delivery policies and procedures into the hospital and its units 
 \cite{Bender:2014}. Importantly, CNL students conducted their Master’s level microsystem change project, prior to the formal intervention implementation time, in the same unit they would be working on as part of the care delivery redesign intervention. This may or may not have influenced the ‘change point’ of the intervention effect, and was therefore considered a good ‘test case’ for modeling purposes. 

The remainder of this paper is organized as follows. First, we present a background of studies on interrupted time series in healthcare. Current statistical methods and their limitations will be discussed. Then our proposed Robust-ITS model is described. Details on the estimation and inference procedure are provided. Followed by an analysis of the impact of Clinical Nurse Leader on patient satisfaction with nurse communication. Parameter estimates are presented and compared to results obtained via traditional ITS methodologies. Lastly, a summary of the Robust-ITS model and a brief description of future work is provided. 

\section{Background}
The traditional ``gold standard" for evidence generation of healthcare 
interventions is the randomized clinical trial (RCT). The theory behind this 
methodology is that potential biases related to patient heterogeneity 
and confounding covariates are evenly dispersed across study groups, 
and thus do not dissimilarly influence treatment effect \cite{Rickles:2009}. 
In statistical terms, the RCT design tests for the difference in outcomes 
between two groups ---- those exposed to the treatment and those not ---- 
completely ignoring underlying variability. However, RCTs have a narrow scope in the care delivery community since it is not feasible, and sometimes not ethical, to randomly assign the intervention.  By design, explanatory RCTs do not and cannot take into account the range of dimensions of patient demographics, variations in health, and overall healthcare complexity \cite{Petticrew:2013}.

Interrupted time series (ITS) offers a rigorous methodology to determine the effectiveness of complex healthcare interventions on outcomes in real world settings, that account for secular changes as part of the analytic process \cite{Taljaard:2014kc, Kontopantelis:2015cn}.  
When RCTs are not feasible or not applicable, ITS is considered the strongest research design in the health policy evaluation literature \cite{Penfold:2013bc}, 
 and are considered rigorous enough for inclusion into Cochrane meta analyses \cite{EPOC:2001tx}.  

\begin{figure}[h!] \centering
	\includegraphics[width=14cm,height=8.5cm]{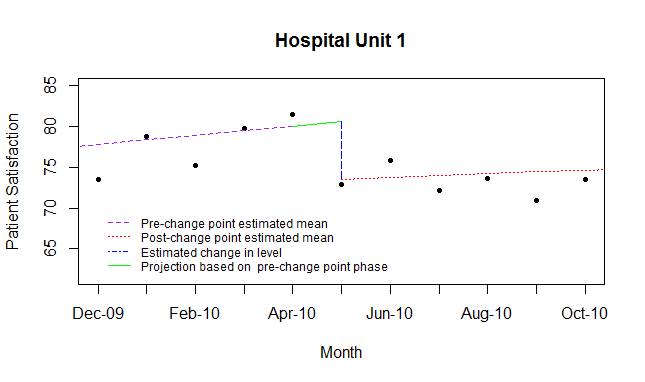} 
	\caption{An example of a segmented regression model fit on patient satisfaction for Unit 1. The plot depicts (1) the segmented regression lines fit to the pre- and post-change point phase, (2) the projection of the mean value at the change point based on the pre-change point regression, and (3) the change in level as defined here. The plot contains data from December 2009 to October 2010, instead of the entire observational period, to clearly illustrate the level change.}
	\label{CIL}
\end{figure}

The metrics adopted by the health policy evaluation literature to assess the effect size of an intervention via ITS are level change and trend change (change in slopes).
A change in outcomes is referred to as a level change and is analogous to the 
difference in mean scores before and after the intervention, with independent data values. The level change is interpreted as the jump between the projected mean based on the pre-change point phase and the estimated mean post-change point.
Our definition of level change is graphically depicted in Figure \ref{CIL}. 
While the level change identifies the size of an intervention's effect, the change in trend quantifies the impact of the intervention on the overall mean. It is necessary to report both level change and change in trend to interpret the results of an ITS study accurately \cite{EPOC:2015ww}. 

\noindent {\bf Limitations of Segmented Regression Approaches.} We first note that  Ramsay et al. (2003) ``demonstrated that ITS designs are often analyzed inappropriately, underpowered, and poorly reported in implementation research". \cite{Ramsay2003} 
We believe, along with many authors in this field, that segmented regression is most effective in analyzing ITS data. Segmented regression may be utilized via standard statistical packages --- such as ITSA in Stata, ETS in SAS, segmentedR in R, etc. ---  however, as we demonstrate in this paper, there are limitations to these current statistical packages. Segmented time series regression, or regression-discontinuity analysis, was first introduced by Quandt (1958) closely followed by Thistlethwaite and Campbell (1960). \cite{quandt1958, Thistlethwaite:1960wy}. Since then, regression-discontinuity analysis has been used in many forms, with many different parameterizations, in health services research (as already described) and other fields, such as economics and education. 

We discuss the potential limitations of the current 
regression modeling approaches, some of which are described in a recent review by  Kontopantelis et al.(2015).\cite{Kontopantelis:2015cn} In particular, regression modeling approaches assume the pre-intervention mean is linear and that the characteristics of the population remain unchanged throughout the study period. Segmented time series regression also assumes that there is a distinct separation between the pre- and post-intervention phases --- by either assuming the time point at which the effect of the intervention initiates (i.e., the change point) is known or removing the set of possible change points ---  and that there are two regressions present, but one overall variance and autocorrelation. These assumptions, if violated, could lead to incorrect inference and interpretation of results. 

The specification of the change point as the time of intervention in segmented regression does not represent the reality that complex interventions may have varied effects and take time to manifest change, and can therefore lead to incorrect measures of the intervention's effect on the system.
Current statistical programs often assume an instantaneous intervention effect --- a change point set to intervention time --- because change point estimation involves optimization over all possible configurations, which challenges computational feasibility. 
Recall, the change point is the time point at which the effect of the intervention initiates.
Prevalent approaches to overcoming this limitation are to remove, or censor, a specific set of time points from the analysis \cite{Penfold:2013bc, Taljaard:2014kc}. 
This censoring not only omits data, but it also potentially biases the parameter estimates, as the study team decides which time points to remove. 

The assumption of a constant correlation structure (variance and autocorrelation) is not necessarily representative of complex interventions, where the system seldom reacts in an isolated way to change, and the intervention is expected to reduce variability in the system. 
In fact, with complex interventions, often the goal is to enhance care processes so that elements become more dependent and consistent over time; theoretically the correlation structure should differ based on changes,  \cite{Cabrieto:2016ba}  such as an intervention. An increase in data dependency and consistency implies a difference in autocorrelation and variability.
Thus, detection of differences in autocorrelation and variances between pre- and post-intervention are critical in evaluating the effectiveness of an intervention.
Table \ref{MComp} highlights a few popular ITS packages, some articles that describe the use of the packages for analysis of ITS data, and the limitations of each method, as already described. 

\singlespacing
\begin{center}
	\begin{table}[H] 
		\small{\begin{tabular}{ |p{2.2cm}|p{4.5cm}|p{4.8cm}|p{3.7cm}| }				
				\hline
				Package & Papers & Advantages/Description & Limitations \\
				\hline
				SAS PROC  AUTOREG   & Penfold \& Zhang (2013), \cite{Penfold:2013bc}  Shardell et al. (2007)\cite{Shardell:2007} & $\bullet$ Estimation and prediction of linear  regression models    &$ \bullet$ No intervention analysis. \\
				\cite{SAS} &   Parienti et al. (2011) \cite{Parienti:2011}  & with autoregressive errors.  $\bullet$ Estimation and testing of general heteroscedasticity  (change in variance). & \\
				\hline
				
				SAS PROC \, \, \, \,   \, \, \, \, ARIMA  \, \, \, \,  \cite{SAS} &Shardell et al. (2007), \cite{Shardell:2007} & $\bullet $ Analyzes and forecasts time series, transfer functions,  and intervention data using ARIMA and ARMA models.   & $\bullet \,$Assumes intervention time is fixed with an immediate effect. \, \, \, \,\,\,\,\,\,\,\, $\bullet $ Assumes one  overall correlation \\
				& &   & structure. \\
				\hline
				
				SAS ETS  \cite{SAS} &  Cable (2001),\cite{Cable:2001} \,\,\,\,\,\, \,\,\,\,\,\,  \,\,\,\,\,\,  Mahamat et al. (2007),\cite{Mahamat:2007}  \,\,\,\,\,\,Aboagye-Sarfo et al. \,\,\,\,\,\, (2015), \cite{Aboagye-Sarfo} &  $\bullet$ Same as the above two entries; SAS PROC ARIMA and SAS PROC  AUTOREG are part of SAS ETS.& $\bullet$ Fixed intervention time with immediate effect. \,\,\,\,\,\,\,\,\,\,\,\,\,\,\,\,\,\,\,\,\,\,\,\,\,\,\,\,\,\,\,\,\,\,\,\,\,\,\,\, $\bullet$ One overall  \\
				&   &  &   correlation structure   \\
				%&& &  \\
				\hline 
				Stata ITSA  &  Linden (2015) \cite{Linden:2015vo}& $\bullet$ Single and multiple group comparisons. \, \, \, \, \, \, \, \, \, \, \, \,\, \, \, \, $\bullet$ Estimates treatment effects for multiple treatment periods. \, \, \, \,\,\,\,\,\,\,\,\,\, \, \, \, \, \, \, \, $\bullet$ Adjust for overall autocorrelation. & $\bullet$ Fixed change point. \,\,\,\,\,\,\,\, $\bullet$ One overarching correlation structure. \\
				\hline
				segmentedR & Muggeo (2012) \cite{Muggeo:2015}& $\bullet$ Estimates piecewise regression models with a fixed number of discontinuities, or interruptions.  & $\bullet$ No modeling of correlation structure. \, \, \, \,\,\,\,\,\,\,\,\, $\bullet$ Assumes data are independent. \\
				\hline			
			\end{tabular}
		}
		\caption{Denotes the limitations and advantages of ITS packages that focus on segmented regression, and gives a few papers in which the packages have either been proposed or utilized. } \label{MComp}
	\end{table}
\end{center}

\doublespacing

\section{The  Robust Interrupted Time Series (Robust-ITS) Model} 
\subsection{Preliminary Analysis}
 Before any formal statistical modeling, the outcome should be plotted against time to illuminate the type of longitudinal mean (linear, quadratic, etc.), seasonality, and the set of plausible change points. The set of possible change points should not be limited to time points solely after the intervention, for aforementioned reasons. If the longitudinal mean is not linear, an adequate transformation may be applied to obtain a linear pattern, or a different segmented regression model appropriate for the pattern present needs to be applied within the ITS design. Seasonality should be accounted for, within the mean, via traditional statistical methods concisely described in Bhaskaran et al. (2013).\cite{Bhaskaran2013} If needed one should apply variance stabilizing transforms to the outcome variable.  For the purposes of illustrating Robust-ITS, the relationship between the outcome and time is assumed linear with no seasonality. 

\subsection{Description of the Robust-ITS Model} \label{Section3} 
One prominent feature of our approach in the Robust-ITS model is the clear distinction between the time of intervention and the change-point.
In Penfold and Zhang (2013), Garey et al. (2008), Ansari et al. (2003),
and many more, the impact of the intervention is assumed to be instantaneous --- that is, the change point is assumed to be the intervention time.\cite{Penfold:2013bc, GareyLai, AnsariGray} Robust-ITS allows us to estimate the time point at which the effect of an intervention initiates. 

The paramount contribution of Robust-ITS is the modeling of the stochastic component separately between the pre- and post-change point phases. The separate modeling allows for two completely different data dependency and variability structures to exist prior to the intervention and post intervention. 
 
Denote $t^*$ as the time point at which the intervention is introduced and $\tau$ as the time point at which the effect of the intervention initiates (the change point). Sometimes it may indeed be true that $t^* = \tau,$ but not necessarily. Often it is entirely possible that
the time of effect of the intervention differs from the time of intervention introduction (i.e., either $\tau > t^*$ or $\tau < t^*$).
Here we develop a data adaptive procedure for estimating $\tau.$ 
There are many change-point detection methods in time series but they often deal only with changes in the mean and variance (not the autocorrelation structure itself), and may not work well in shorter time series \cite{davis2006, Ombao:2015}.

Define $Y_t$ as the outcome of interest at time $t$; for 
example, $Y_t$ may be patient satisfaction at a particular hospital unit during time $t$. 
The general regression is defined as
$$ Y_{t} = \mu_{t} + \epsilon_{t}, $$ where
$\mu_{t}$ is the mean and $\epsilon_{t}$ is the 
stochastic process.
The mean component, $\mu_t$, characterizes the mean of the outcome for the pre-intervention and post-intervention phases.
The stochastic process, $\epsilon_t$, accounts for the outcome variability and correlation. 
In the following discussion we define the mean and stochastic components for the Robust-ITS model.
A note on the length of the time series needed to carry out the Robust-ITS analysis is provided in the Appendix. 

\subsubsection{The Pre- and Post-intervention Mean}
At the first stage of modeling the emphasis is on the mean, 
\begin{equation} \label{EQ3}
\mu_{t} = \left\{
\begin{array}{lr}
\beta_{0} + \beta_{1} \, t,  &  \,  t <  \tau \\
( \beta_{0} + \delta) \, + (\beta_{1} + \Delta) t,  &  t \geq \tau ,
\end{array}
\right.
\end{equation}
where the parameters are estimated using ordinary least squares.
The parameters in $\mu_t$ are: (1.)  $\beta_{0},$ the intercept of the mean prior to the change point; (2.) $\beta_{1},$ the slope of the outcome prior to the change point; (3.) $\beta_0 + \delta,$ the intercept of the post-intervention phase; and (4.) $\beta_1 + \Delta,$ the slope of the post-intervention phase.

\vspace{0.17in}
\noindent {\bf Remark.} (1.) The difference between the pre-change point and post-change point intercept is $\delta$;  (2.) The difference between the pre- and post-change point slopes is $\Delta$; (3.) the difference in the mean level (pre- minus post-intervention) is $-\delta - \Delta \tau,$ the level change. 
\vspace{0.1in}

Recall, difference in mean level (or  level change) is one of the two metrics in health policy evaluation literature used to measure the effect size of an intervention. Formally, the level change is defined as the 
difference at the change point time $\tau$ between the extrapolated pre-intervention mean 
level and the observed intervention mean level, as is depicted in Figure \ref{CIL}.

Rather than {\it impose} or {\it assume} the actual onset of  the change, the Robust-ITS model actually {\it estimates} the change point $\tau$ in a data-driven manner using the likelihood approach. From a set of candidate change points (set by the researcher), the
procedure estimates the parameters via ordinary least squares for each possible $\tau,$ and selects the $\tau$, and its corresponding parameters, that maximize the likelihood. 

Denote the length of the time series as $T$ and let $\theta =  [ \beta_0, \beta_1, \delta, \Delta, \sigma_1^2, \sigma_2^2 ]',$ with $\sigma_1^2$ and $\sigma_2^2$ defined as the variances prior to and post change point respectively. As described in section 2, one goal of interventions is to decrease variability, which leads to creating a more consistent outcome.
We therefore include separate variance parameters for the pre- and post-change point phases, to allow for a change in data variability.

Let $q$ be a candidate change point in the set of possible change points $Q,$ where $Q = \{ t^* - m, \dots, t^*, \dots t^* + k \}$ for positive integer values of $m$ and $k$ set by the researcher. For each candidate change-point $q$ we derive the likelihood function:
\begin{align*}
& L(\theta  \vert q) = \,\,   \Big( \frac{1}{\sqrt{2 \pi \sigma_1^2}} \Big)^{q-1} \text{exp} \Big( - \frac{1}{2 \sigma_1^2} \sum_{t=1}^{q-1} \big[Y_t - (\beta_0 + \beta_1 t) \big] ^2 \Big) \, \times
 \\
& \,\,\,\,\,\,\,\,\,\,\,\,\,\,\,\,\,\, \,\,\,\,\,\,\,\,\,\,\, \Big( \frac{1}{\sqrt{2 \pi \sigma_2^2}} \Big)^{T - (q-1)}  \text{exp} \Big( - \frac{1}{2 \sigma_2^2} \sum_{t=q}^{T} \big[Y_t - (\{\beta_0 + \delta\} + \{\beta_1 + \Delta \}t) \big]^2 \Big). 
\end{align*}
Define $ L_{\max} (q)  =  \max_{\theta} L(\theta, \vert q),$ then the estimated change point is
 $\hat{\tau} =  \arg \max_{ q \in Q} \, L_{\max} (q).$

The estimates of the intercept and slope for each phase are obtained as in segmented regression; equivalent to estimating the slope and intercept separately for the pre- and post-change point phases as in simple linear regression. The ordinary least squares (OLS) estimates for the parameters in $\theta$ are provided in the Appendix.
The estimates for $\sigma_1^2$ and $\sigma_2^2$ depend on the stochastic process, and are given for an AR(1) process also in the Appendix.

The presence of $\tau$ does not restrict the model to a fixed interruption with an instantaneous effect, and allows the design matrix and estimates to transform based on the information the data provides.
This flexibility of the model can be helpful in minimizing misleading results from an assumed change-point.  
 
\subsubsection{Stochastic Properties Pre- and Post- Change Point}
The stochastic component, $\epsilon_t$, captures the correlation structure of the outcome variable across time, and may change as a result of the intervention. Here, we shall develop a formal test for the difference in the correlation structure for pre- and post-intervention phases. 
  
We use the ARIMA process to model the stochastic component, $\epsilon_t = Y_t - \mu_t$. Since the mean function $\mu_t$ is not known (we only have its estimate, $\hat{\mu_t}$), the stochastic component is not directly observed. In place of $\epsilon_t$, we use the residuals, $R_t = Y_t - \hat{\mu_t}$, where $\hat{\mu_t}$ is the estimate of $\mu_t$ obtained as described in stage one. In order to use the ARIMA processes, residuals must exhibit stationary behavior, that is, the mean and variance of the residuals must be relatively constant. If the mean is not misspecified, then the residuals should be fluctuating around zero without any patterns. Moreover, the residuals should be stationary within each of the pre- and post-intervention phases.  \cite{Shumway:2011}

Autoregressive conditional heteroscedasticity models may be used when the data is non-stationary; they can model the stochastic component in each phase and for the entire observational period when the variance and/or data dependency is non-constant. 
For our patient satisfaction data it is reasonable to assume stationarity within each phase, and hence we proceed with the assumption of stationarity. 
See Shumway and Stoffer (2011), Granger and Newbold (2014), and Bollerslev (1988) in for more details on autoregressive conditional heteroscedasticity models.\cite{Shumway:2011, granger2014, bollerslev1988}  
  
Due to the impact of the intervention, the stochastic process $\epsilon_t$ pre-intervention might differ from the process post-intervention. 
 That is, $\epsilon_{t}$ for $t \in \{1, \dots, \hat{\tau} -1 \}$ may be a different stochastic process than $\epsilon_{t}$ for $t \in \{ \hat{\tau}, \dots, T\}.$ Hence, the autocorrelation and variance might differ pre- and post- change point. Now, the stationarity requirement is satisfied if the variance, mean, and autocorrelation are constant within each stochastic process, not constant across all time points as before. 
 
The ARIMA parameters are estimated by maximizing the conditional likelihood, and are given for the subsequent example of an autoregressive model with a lag of one. It is of most importance to understand that the lag used for the autocorrelation modeling is not an indicator of when the intervention takes effect, but instead it models overall data dependency; $\tau$ dictates when the intervention affects the outcome variable. 
 
 \vspace{0.10in}
 
 \noindent {\bf Example.} A special case of a stochastic process is the first order auto-regressive [AR(1)] model: 
 \begin{equation} \label{AR1}
 R_{t} = \left\{
 \begin{array}{lrr}
 \phi_1 \, R_{t-1} + e_t^{1},  &  & \,\, \, \,  \, \, \,\, \, \,  \, \, \,  \, \, \, \, \, \, \, \, \,  \, \, \, 1 < t < \hat{\tau}\\
  \phi_2  \, R_{t-1} + e_t^{2}  &   &  \hat{\tau} < t \leq  T. 
 \end{array}
 \right.
 \end{equation}
 To ensure causality in the time series sense, both $\phi_1$ and $\phi_2$ lie in the interval $(-1, 1).$
 Note, $\phi_1$, the auto-regressive coefficient prior to the change point, is directly associated with the correlation between two time points; $\phi_1$ is the correlation between time point $t$ and $t+1$ where $t$ and $t+1$ belong to the pre-change point phase ($t, \text{and  } t+1 \in \{ 1, \dots, \hat{\tau} -1 \}$),  and $\phi_1 ^{\vert h \vert }$ is the correlation between two time points $h$ time periods away (say $t$ and $t+h$ both in the pre-change point phase, $\{ 1, \dots, \hat{\tau} -1 \}$). The auto-regressive coefficient post change point, $\phi_2$ has a similar interpretation. The error terms of  model \ref{AR1} are white noise, $e_t^j \stackrel{iid}{\sim} N(0, \sigma^2_j)$ for $j \in \{1, 2\}.$
   
 The variance and auto-regressive coefficients in the AR(1) setting can be estimated by maximizing the conditional likelihood. The estimates are functions of the residuals $R_t$ and the residuals of the residuals $W_t$, and are provided in the Appendix. 
 
 To determine whether the stochastic process differs as a result of the change point, we test the hypothesis that $\nu \equiv \phi_2 - \phi_1$ equals zero. This can be tested by either estimating $\nu$ directly or by conducting an F-test for nested models. The F-test for nested models for this AR(1) scenario is described in the Appendix.

\vspace{0.17in}
\noindent \textbf{Remark.} Once the $\epsilon_t$ are appropriately modeled, the OLS estimates in equations \ref{MLE0} -\ref{MLE3} will need to be re-estimated to produce the generalized least squares estimates. If an AR(1) is fit to the overall stochastic process (across the change point), the beta parameters should be re-estimated without the first time point; that is, for $t$ in$\{ 2, \dots, T\}.$ If different AR(1) processes are fit pre- and post-intervention, then the mean prior to the intervention should be re-estimated using $t$ in$\{ 2, \dots, \tau -1 \},$ and the mean post intervention re-estimated using $t$ in$\{ \tau + 1, \dots, T\}.$ The summation limits in equations \ref{MLE0} -\ref{MLE3}  would therefore change. 
 
\subsubsection{Pre- and Post-Intervention Variance Comparison  }
 
 For stochastic processes in which both pre- and post-change point phases are adequately modeled by the ARIMA processes (residuals not behaving as white noise in either phase), the variances may not be easily, if at all, compared. The variances in each phase can be estimated but not statistically compared, due to the dependency of the data. 
 
 If there is no autocorrelation (or dependence) then the OLS estimates are sufficient. 
 Nevertheless, the variance may not be the same pre- and post-change point. In situations where there is no statistically significant autocorrelation, the variances may be compared via an F-test. Using $\tau$ we can determine how many observations we have prior to and post change point, subtracting three (one for each parameter we estimate) from those values gives the degrees of freedom. For example, suppose there are 25 and 35 time points before and after the change point respectively, and that the estimated variances are $s_1$ and $s_2$ respectively. Then the F-statistic is $\frac{s_1}{s_2}$, and under the null hypothesis (assuming the variances are equal)  distributed $F_{22, 32}.$ 
 
\section{Robust-ITS Analysis of the Intervention Effect on Patient Satisfaction }

Patient satisfaction is modeled in four hospital units, respectively labeled: Unit 1, Unit 2, Unit 3, and Unit 4. It is crucial to note that the outcome is a percentage, and so, restricted to lie between 0 and 100. The restriction on the outcome has imperative consequences: the time series must reach a plateau regardless of intervention introduction. The nature of the outcome must be kept in mind when interpreting the results from the analysis. 

The means of the four time series were modeled as in equation \ref{EQ3}; the resulting parameter estimates are given in Table \ref{Res}. The relationship between the formal intervention implementation time and the change point for the four units is illuminated in Table \ref{Res} and Table \ref{CPTable}, which show the effect of the intervention is not necessarily instantaneous. In fact, Table \ref{Res} and Table \ref{CPTable} suggest the intervention had an anticipatory effect in three of the four units of interest. The preemptive effect is in concordance with the structure of the CNL integrated care delivery intervention, because of the CNL student inclusion into their respective units 6 months prior to the formal introduction.
In Units 1, 2 and 3 the estimated change points occur respectively in May 2010, January 2010, and February 2010, suggesting CNL students could have implemented the new care delivery prior to July 2010. This relationship indicates the time of change in patient satisfaction associated with the intervention may be at the mercy of CNL student behavior. 

\singlespacing

\begin{table}[H] 
	\centering
	\small{ \begin{tabular}{ |p{3cm}|p{3cm}|p{3cm}|p{3cm}|p{3cm}| }
			\hline
			\hline
			\multicolumn{5}{|c|}{Patient Satisfaction} \\
			\hline
			Parameters & Unit 1 & Unit 2 & Unit 3 & Unit 4 \\
			\hline
			Intercept Pre & 64.32**  & 64.67** 
			& 68.31**   & 77.21** \\
			Change Point 	&  (61.76, 66.88) &  (60.07, 69.27) 
			&  (64.14, 72.49)  &  (74.99, 79.54) \\
			\hline
			Intercept Post & 67.21**  & 71.79**  
			& 71.51**  & 77.42**  \\
			Change Point & (61.86, 72.57)  &  (66.10, 77.49) 
			&  (64.01, 79.01) &  (70.24, 84.60) \\
			\hline
			Change in &  2.89 
			& 7.12  & 3.20  & 0.15 \\
			Intercepts, $\hat{\delta}$ &  (-2.91, 8.70)
			&  (-0.03, 14.28)  & (-5.22, 11.61) & (-7.20, 7.51) \\
			\hline		
			Change in level, & 7.00**
			& -2.77 & 3.50  & 5.40** \\
			$-\hat{\delta} - \hat{\Delta}\hat{\tau}$ &  (3.75, 10.25)
			&  (-7.92, 2.38) &  (-1.72, 8.72) &  (2.01, 8.78)\\
			\hline
			Slope Pre  & 0.56**    & 0.24    & 
			0.35*   & 0.28**   \\	
			Change Point &  (0.41, 0.71)   &  (-0.08, 0.56)   & 
			(0.07, 0.63)   &  (0.15, 0.40)  \\	
			\hline
			Slope Post  & 0.22**  & 0.07  & 
			0.09  & 0.10\\
			Change Point & (0.10, 0.34)  &  (-0.06, 0.20)  & 
			(-0.07, 0.26 )  &  (-0.06, 0.25) \\
			\hline
			Change in Slope,  &  -0.34** 
			& -0.17   & -0.26   & -0.18  \\
			$\hat{\Delta}$ & (-0.53, -0.15)  
			&  (-0.51, 0.16)  &  (-0.58, 0.06)  &  (-0.38, 0.02) \\
			\hline
			Delay in Effect of  &  & & &\\
			Intervention, &  -3
			& -6 & -5 & 0\\
			$\hat{\tau} - t^*$ & & & & \\
			\hline
	\end{tabular} }
	\caption{Provides 95\% confidence intervals and estimates of the mean parameters for average patient satisfaction of Unit 1, Unit 2, Unit 3, and Unit 4. Since $\tau$ is discrete, only an estimate is given, no confidence interval. The asterisk, *, denotes statistical significance at the $\alpha=.05$ level. } \label{Res}
\end{table}

\begin{table}[H]
	\centering
	\small{ \begin{tabular}{|p{4.2cm}|p{2.4cm}|p{2.4cm}|p{2.6cm}|p{2.4cm}| }
			\hline
			\hline
			\multicolumn{5}{|c|}{Patient Satisfaction} \\
			\hline
			& Unit 1 & Unit 2 & Unit 3 & Unit 4 \\
			\hline
			Time of Intervention  & Month 31, & Month 31,  & Month 31, & Month 31,  \\
			Implementation & July 2010 &  July 2010 &  July 2010 &  July 2010 \\
			\hline
			Estimated Change   & Month 29, & Month 25,  & Month 26, & Month 31,  \\	
			Point, $\tau$	&  May 2010 &  January 2010 &  February 2010 &  July 2010 \\		
			\hline
	\end{tabular} }
	\caption{Gives the formal time of intervention implementation and the estimated time at which the effect of the intervention initiates, the change point.} \label{CPTable}
\end{table}

Table \ref{Res} depicts the differences in estimated means prior to and post change point, with the most informative rows of Table \ref{Res} corresponding to the two standardized effect sizes: change in level and change in slopes. The level change is positive and statistically significant (at the $\alpha = 0.01$ level) for Unit 1 and Unit 4, indicating that the mean drops at the change point and that the drop statistically differs from zero. Thus, the CNL integrated care delivery initially is associated with a statistically significant drop of patient satisfaction in Unit 1 and Unit 4. The estimated trend change or change in slopes is negative for each unit, although statistically significant (at the $\alpha = 0.01$ level) for Unit 1 only. 

The slope decreases after the estimated change point in Unit 1, implying a more flattened out mean post-change point. Therefore, the CNL implementation may be associated with a flatter mean across time in Unit 1; i.e., for every one month increase in time, there is a smaller estimated increase in patient satisfaction in the post-change point phase as compared to the pre-change point phase. However, this artifact may be present because the maximum value of the outcome variable is 100.
We may be seeing some asymptote effect instead of capturing the effect of the intervention on the trend (slope).

For Units 2, 3, and 4 the estimated slope does not statistically change after the estimated change point; for Units 2 and 3 the estimated level change is also not statistically significant; and, the estimated change in intercepts is not statistically significant for any of the units. Hence for Units 2 and 3, the intervention does not seem to be associated with a change in the estimated patient satisfaction. 
The CNL integrated care delivery is associated with some outcome modification (either in the intercept, level change, slope change, or a combination) in Unit 1 and Unit 4. 

The pre- and post-change point regressions of the four units are plotted in Figure \ref{UnitTS}. Figure \ref{UnitTS} depicts that the change point occurs prior to the formal intervention time for Unit 1, 2 and 3, but is equivalent to the formal intervention time for Unit 4. The estimated mean post-change point seems to flatten out in all units, and
the change in level appears sizable for Unit 1 and Unit 2. Figure \ref{UnitTS} illustrates results in concurrence with those of Table \ref{Res}. 

Figure \ref{Resplots} in the Appendix provides the studentized residuals after modeling the mean. The residuals seem well behaved and mostly contained between the rule of thumb $\pm 2$ and completely contained between $\pm 3$. The residuals do not exhibit any severe patterns, and thus suggest Robust-ITS models the mean patient satisfaction of all units adequately. Moreover, Figure \ref{ACFplots} provides the autocorrelation function (ACF) of the residuals. The ACF plots interpreted as Shumway and Stoffer (2011) act as white noise, implying that the data do not exhibit autocorrelation. \cite{Shumway:2011}

Modeling the mean by equation \ref{EQ3} in stage one is sufficient because the residuals act as white noise. Nevertheless, we model the residuals pre- and post-change point with an AR(1) process separately, to provide complete information. The estimates and 95\% confidence intervals of the autoregressive parameters and their difference is given in Table \ref{Table4}, along with the estimated variance prior to and post change point and their comparison. Both $\hat{\phi_1}$ and $\hat{\phi_2}$ do not statistically differ from zero in any of the four units, supporting our claim that the residuals act as white noise. There is no data dependency apparent in either the pre- and post-change point phases. 
The difference of the two autoregressive parameters, $\hat{\phi_2} - \hat{\phi_1},$ also do not statistically differ from zero in the four units.

\begin{table}[!htb]
	\centering
	\small{ \begin{tabular}{|p{4.2cm}|p{2.4cm}|p{2.4cm}|p{2.6cm}|p{2.4cm}| }
			\hline
			\hline
			\multicolumn{5}{|c|}{Patient Satisfaction} \\			\hline
			Parameters & Unit 1 & Unit 2 & Unit 3 & Unit 4 \\
			\hline
			
			AR(1) Coefficient Pre &  -0.056  & -0.191 &  0.078 & -0.271 \\
			Change Point, $\hat{\phi_1}$ & (-0.460, 0.348) & (-0.624, 0.241) &  (-0.377, 0.534) & (-0.647, 0.105) \\
			\hline
			
			AR(1) Coefficient Post & -0.354 & 0.055 & 0.088 & -0.044 \\
			Change Point, $\hat{\phi_2}$ &  (-0.713, 0.004) & (-0.266, 0.376) & (-0.264, 0.440) & (-0.401, 0.392) \\
			\hline	
			
			Difference in AR(1)  & -0.299 & 0.246 & 0.010 & 0.267 \\
			Coefficients, $\hat{\phi_2} - \hat{\phi_1}$ & (-0.826, 0.229) & (-0.278, 0.770) & (-0.551, 0.570) &(-0.267, 0.801) \\
			\hline	
			
			Variance Pre Change Point, $\hat{\sigma^2}_{1:(\hat{\tau} -1)}$ & 10.259 &  26.474 & 23.412 & 8.127  \\
			\hline	
			
			Variance Post Change Point, $\hat{\sigma^2}_{\hat{\tau}: T}$	&  7.976
			& 13.511 & 23.965 & 12.649 \\
			\hline		
			
			Variance Comparison & 1.286 & 1.959  & 0.977 & 0.643 \\
			F-statistic (p-value) & (0.248) & (0.035) & (0.516) & (0.88) \\
			\hline
	\end{tabular} }
	\caption{Gives (a.) estimates and 95\% confidence intervals of the AR(1) coefficients pre and post change point, and of the estimated increase in the AR(1) coefficient post-change point; (b.) the estimated variances and (c.) the F-statistic and p-value corresponding to the comparison of the pre and post change point variances, for patient satisfaction with effective nurse communication. } \label{Table4}
\end{table}

Because there is no correlation present and the stochastic component is adequately modeled by white noise (indicating independent data), there is valuable information obtained from $\epsilon_t$ via the variance; the variances are compared using an F-test. The estimates of the variances are smaller post-change point for Unit 1 and Unit 2, and larger for Unit 3 and Unit 4. Nonetheless, we cannot conclude that the variance differs between the two phases for Unit 1, Unit 3 and Unit 4. For Unit 2 the variance post-change point is statistically (at the $\alpha = 0.05$ level) smaller than the variance pre-change point. Therefore, patient satisfaction in Unit 2 is more predictable after the introduction of CNL integrated care delivery. A more predictable outcome, less extremely unsatisfied and satisfied patients, signifies a more controlled environment. This is a positive result of the intervention since there will be better quality control on the fluctuations of the patient outcomes and more consistency as a result of the intervention. 

\vspace{.3cm}
\noindent {\bf Comparing Robust-ITS to Segmented Regression}
We compare the standardized effect sizes between Robust-ITS and segmented regression (both with an assumed change point and the set of possible change points removed) in Table \ref{Models}. 
The aim of Table \ref{Models} is to illustrate that the estimates of level change and trend change differ based on the type of model selected. Indeed, the estimates of level change and trend change across the 3 models differs for each of the fours units.
Segmented regression --- with an assumed change point or the set of possible change points removed --- may provide results that are statistically significant, or not statistically significant, in cases where the opposite is true when considering anticipatory or delayed intervention effects. Moreover, the two segmented regression methods may also provide opposing results.

\begin{table}[!htb] 
	\centering
	\small{ \begin{tabular}{ |p{1.5cm}|p{2.4cm}|p{2.4cm}|p{2.2cm}|p{2.2cm}|p{2.4cm}|p{2.2cm}| }
			\hline
			\hline
			\multicolumn{7}{|c|}{ Patient Satisfaction } \\
			\hline
			Unit &  \multicolumn{3}{|c|}{Change in Level}  & \multicolumn{3}{|c|}{Change in Trend (slope)}  \\	
			&  \multicolumn{3}{|c|}{-$\hat{\delta} - \hat{\Delta} \hat{\tau}$}  & \multicolumn{3}{|c|}{$\hat{\delta}$}  \\
			\hline 
			& Segmented Regression+  & Segmented Regression++ & Robust-ITS & Segmented Regression+  & Segmented Regression++ & Robust-ITS \\
			\hline
			Unit 1 & 6.04&  5.7  & 7 & 
			-0.41 & -0.25 &  -0.34  \\
			& (1.14, 10.94) &  (2.32, 9.07)  &  (3.75, 10.25)   & 
			(-0.68, -0.15) &  (-0.45, -0.06) & (-0.53, -0.15)   \\				
			& 0.02* &  0.00** &    0.0** & 
			0.00** & 0.01*&  0.00** \\
			\hline
			
			Unit 2 &  -4.4 & -1.8   &  -2.77
			&-0.24 &  -0.23  &  -0.17  \\
			& (-11.01, 2.21) & (-6.38, 2.79) &   (-7.92, 2.38) 
			& (-0.61, 0.14) & (-0.50, 0.04)  &    (-0.51, 0.16) \\
			& 0.19 & 0.44  &  0.29 
			& 0.21 & 0.10 &  0.30\\			
			\hline
			
			Unit 3 &  0.94 & 0.59  & 3.50  &
			-0.21 & -0.16  &  -0.26 \\
			&  (-6.64, 8.53) &	(-4.86, 6.03)    & (-1.72, 8.72) &
			(-0.60, 0.18) & (-0.46, 0.14)  &   (-0.58, 0.06)  \\
			&   0.8 & 0.83  &  0.18 
			& 0.28 & 0.28 &  0.11 \\
			\hline					
			Unit 4& 5.89 & 5.40   &  5.40   &
			-0.28 & -0.18  & -0.18   \\
			& (0.64, 11.14) & (1.99, 8.80)  &  (2.01, 8.78)  &
			(-0.56, -0.01) & (-0.38, 0.02) &  (-0.38, 0.02)   \\
			& 0.03 & 0.00 &  0.00  &
			0.05 & 0.07 & 0.07  \\
			\hline 
	\end{tabular} }
	\caption{Provides approximate 95\% confidence intervals for the level change and trend change of (1.) segmented regression with the phase-in period removed, denoted by +, (2.) segmented regression with an assumed change point, denoted by ++, and (3.) Robust-ITS, for patient satisfaction. The first row within each unit corresponds to the estimate, the second to the confidence interval, and the third to the p-value. Note, one asterisk, *, denotes significance at the $\alpha =0.05$ level, and two asterisks, **, denotes significance at the $\alpha =0.01$ level.  } \label{Models}
\end{table}	

It is important to note that there are many model specifications used for segmented regression. Two of the main models used for segmented regression in the ITS and healthcare literature are discussed and shown to be equivalent in the Appendix. The segmented regression models are discussed under the assumption that the change point is assumed. Nonetheless, the two main segmented regression models are also equivalent when the set of possible change points are removed. 

The true model comparisons are provided in Table \ref{PSE}, intended to compare the adequacy of Robust-ITS and segmented regression. Mean squared error (MSE) --- the estimate of sum of squared errors, which measures the square of the deviations from the estimate mean, divided by the degrees of freedom --- is provided in Table \ref{Models} for Robust-ITS, segmented regression with an assumed change point, and segmented regression with the set of possible change points removed. Robust-ITS has the smallest MSE and so provides the best estimate for the mean of patient satisfaction, suggesting that Robust-ITS models the data better than either of the traditional segmented regressions (with an assumed change point, or the set of possible change points removed).

\begin{table}[!htb]
	\centering
	\small{ \begin{tabular}{ |p{1.4cm}|p{4.5cm}|p{4.5cm}|p{2cm}|  }
			\hline
			\hline
			\multicolumn{4}{|c|}{Mean Squared Error} \\
			\hline
			Unit
			& Segmented Regression+  & Segmented Regression++ & Robust-ITS \\
			\hline
			Unit 1 &   192.42 & 190.99 & 171.99   \\
			\hline	
			Unit 2
			&   428.34 & 416.50  & 406.85  \\
			\hline			
			Unit 3 &
			471.80 & 471.93   &  459.16  \\
			\hline						
			Unit 4  & 
			181.44  & 161.11  & 161.11  \\
			\hline			
			
	\end{tabular} }
	\caption{Provides the mean squared error (MSE),  with order of magnitude $10^{-5},$ of (1.) segmented regression with the phase-in period removed, denoted by +, (2.) segmented regression with an assumed change point, denoted by ++, and (3.) Robust-ITS, for patient satisfaction. Mean squared error is the estimate of sum of squared errors, which measures the square of the errors or deviations, divided by the degrees of freedom.
		A lower value of MSE for a model, suggests a more adequate fit.}
	\label{PSE}
\end{table}

\begin{figure}[h!] \centering
	\includegraphics[width=16cm,height=9.25cm]{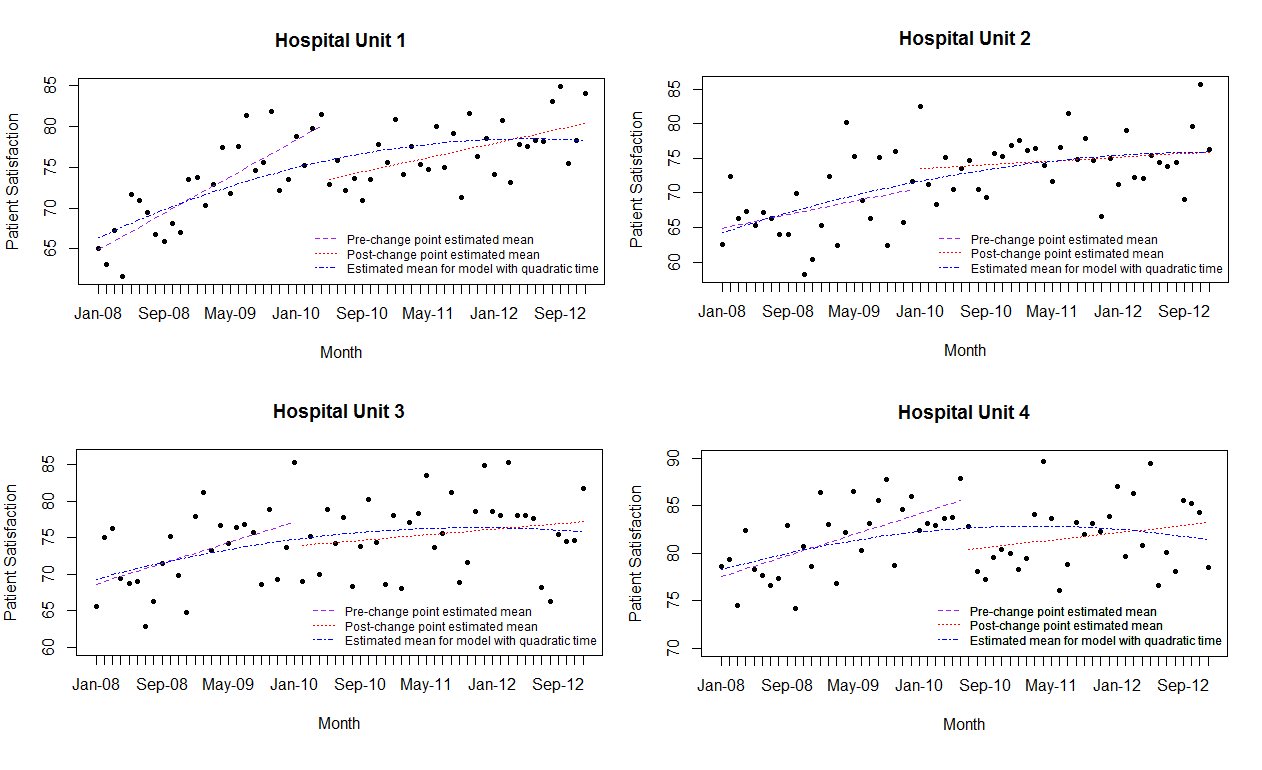} 
	\caption{Plots of patient satisfaction within each of the four units, along with the estimated means obtained by Robust-ITS and a model with quadratic time as a predictor.}
\label{Quad}
\end{figure}

\vspace{.3cm}
\noindent {\bf Comparing Robust-ITS to a Quadratic Model with No Change Point}
We further compare Robust-ITS to a non-change point model with quadratic time as a predictor for completeness.
The model for the mean of patient satisfaction of a given unit with quadratic time as a predictor is 
\begin{equation} \label{mQuad}
\mu_{t}= \beta_0 + \beta_1 \, t + \beta_2 \, t^2  \, \, \, \, \text{for} \, \, t \in \{ 1, \dots, 60\}.
\end{equation}
The estimated patient satisfaction mean curves for both Robust-ITS and model \ref{mQuad} are plotted in Figure \ref{Quad} by unit. The parameter associated with quadratic time $\beta_2$ is only statistically significant, at the $\alpha =0.05$ level, for Unit 1. Including quadratic time as a predictor is not necessary for Unit 2, Unit 3, and Unit 4, since we cannot conclude that $\beta_2$ differs from zero.  Adding quadratic time as a predictor is useful in Unit 1 because at the $\alpha =0.05$ level $\beta_2$ differs from zero. 

Nevertheless as shown in Table \ref{QuadComp}, the MSE (estimate of the sum of squared errors divided by the degrees of freedom) for Robust-ITS is smaller than the MSE of model \ref{mQuad} in all units, indicating Robust-ITS fits the data better in all units. 
Additionally, model \ref{mQuad} assumes a continuous decline after obtaining the maximum. Suggesting model \ref{mQuad} will produce a poor patient satisfaction estimate post maximum. 

\begin{table}[!htb]
	\centering
	\small{ \begin{tabular}{ |p{1.4cm}|P{9cm}|P{2cm}|  }
			\hline
			\hline
			\multicolumn{3}{|c|}{Mean Squared Error} \\
			\hline
			Unit & Model with Quadratic Time and No Change Point & Robust-ITS \\
			\hline
			Unit 1 &   224.91 &  171.99     \\
			\hline	
			Unit 2
			&   417.77  &  406.85   \\
			\hline			
			Unit 3 &
			460.36 &  459.16    \\
			\hline						
			Unit 4  & 
			184.62 &   161.11    \\
			\hline			
			
	\end{tabular} }
	\caption{Provides the mean squared error (MSE), with order of magnitude $10^{-5}$, for Robust-ITS and for the non-change point model with quadratic time as a predictor, for patient satisfaction. Mean squared error is the estimate of sum of squared errors, which measures the square of the errors or deviations, divided by the degrees of freedom.
		A lower value of MSE for a model, suggests a more adequate fit.}
	\label{QuadComp}
\end{table}

\section{Conclusion and Future Work}

There are two main stages that compose Robust-ITS. The first is modeling the mean and the second is modeling the stochastic component. In both stages Robust-ITS tests for a change in the outcome due to the intervention.
To the best of our knowledge, in the ITS literature comparing and testing for a difference in the stochastic component --- a change in autocorrelation and/or variance for the AR(1) case --- has not been considered. 

In the first stage, a set of plausible change points must be established based on the scientific question at hand. Then based on the set of possible change points, Robust-ITS estimates the mean parameters via ordinary least squares and chooses the change point whose parameter estimates maximize the likelihood. In the second stage, the residuals obtained by modeling the mean in the first stage are used to examine and determine the structure of the stochastic process. If the residuals act as white noise, (1) there is no correlation present, (2) the variances before and after the estimated change point are compared by an F-test, and (3) the outcome of interest is adequately modeled by the mean from stage one. Otherwise, an ARIMA process is fit on the residuals pre- and post-change point, separately. From the ARIMA process, estimates of the correlation and variance are obtained via conditional likelihood methods. The correlation estimates are compared to determine if the stochastic process differs as a result of the change point, but the variances are not compared. 

The patient satisfaction and CNL integrated care delivery analysis elucidated that the assumed change point is not always assumed adequately. Following the traditional segmented regression analysis we would have set the change point at the same value for all units, and assumed it was equal to the formal intervention time. We estimated the change point corresponding to CNL integrated care delivery prior to the formal intervention time for three units, and the estimated change point value varied based on unit.

In two of the four units, the CNL integrated care delivery introduction was associated to a change in the mean patient satisfaction. Even though the change in mean patient satisfaction was not necessarily positive, it depicted a mean that continued towards 100\%. The lack of affirmation for the CNL integrated care delivery may stem from the outcome definition as a percentage and an average. The percentage quality of patient satisfaction limits the values the outcome may take on, and thus creates an asymptote effect for units that were already doing well. The averaging across seven patient satisfaction indicators may cancel out improvements in some indicators with regressions in others.

The estimates of the autocorrelation coefficients pre- and post-change point, although not statistically significant, differed by approximately 0.25 for Unit 1, 2, and 4. Since the autocorrelation was not statistically significant, the variances pre- and post-change point were compared. For Unit 2 the variance post-change point was significantly smaller than the variance pre-change point. This is a positive result of the CNL integrated care delivery, since there will may better quality control of patient satisfaction fluctuations due to the CNL intervention.

Comparing Robust-ITS with traditional ITS modeling illustrates how allowing for a variable change point results in a better fit with regards to MSE. The ability to easily assess the effect of the intervention on the correlation structure, and to conduct variance comparisons when correlation is not present, allows for clearer inference on the possible effect of an intervention. 

\begin{figure}[h!] \centering
	\includegraphics[width=16.5cm,height=11cm]{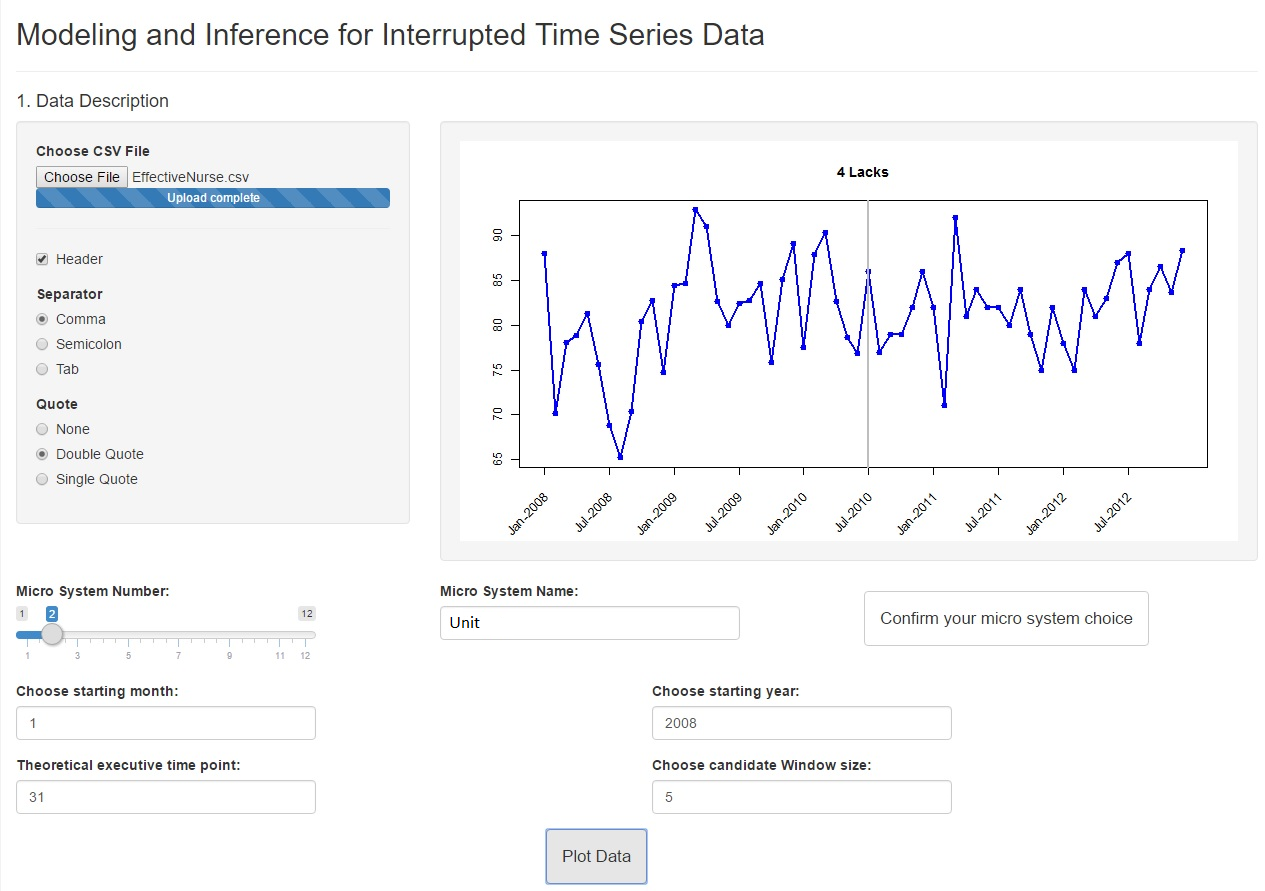} 
	\caption{The Robust-ITS toolbox in R Shiny (by Ngo, Hu, Cruz, Bender, and Ombao), 
		an interactive toolbox in which the user (1.) may upload their own data in a \textit{.csv} file; (2.) provides basic information of the data --- the toolbox requires the user to input the `theoretical executive time point (TET)' (formal time of intervention),  `candidate before TET' (smallest value of the set of possible change points), `candidate after TET' (largest value of the set of possible change points), `starting month' (the month at which data collection began), and `starting year' (year at which data collection began); (3.) views the output plots (after pressing the button labeled `Analyze Data') of the fitted data, the log-likelihood at possible change points, residuals, and acf plots to determine the lag of the stochastic process; (4.) views the estimates, along with their p-values and standard errors, for both the mean and stochastic processes. }
	\label{RShin}
\end{figure}

Our group has developed the Robust-ITS toolbox in R Shiny (see Figure \ref{RShin}) that executes the methodology described here. The toolbox and its manual (in a PDF document) are located respectively at  {\color{blue} \href{http://bit.ly/Robust_ITS}{Robust-ITS}} and {\color{blue} \href{https://www.dropbox.com/s/te5gy1cojtovevo/Manual3.pdf?dl=1}{Manual}}. It is crucial to note that the methodology implemented in the toolbox is the methodology proposed here. Hence, any use of the toolbox should result in the citation of this paper. 
The Robust-ITS toolbox is interactive, and provides the user with graphical displays, estimates and inference on testing for differences between the pre- and post-intervention means, correlation, and variance. 

The current status of the model is only for continuous-valued outcomes. We are now in the process of expanding this to include counts and rates data (e.g., infection rates, counts of accidental falls, etc). Currently, our focus is on developing a segmented regression analog of generalized linear models for count and binary outcomes.
Our aim is to expand the current toolbox, or produce a new toolbox, that will appropriately model count and binary outcomes in a user friendly manner.

\doublespacing				
\bibliographystyle{unsrt}
\bibliography{ITS}	

\begin{thebibliography}{10}

\bibitem{InstituteofMedicineUSRoundtableonEvidenceBasedMedicine:2009fu}
Jandoc R, Burden AM, Mamdani M, Lévesque LE, and Cadarette SM.
\newblock {\em Leadership Commitments to Improve Value in Health Care: Finding
  Common Ground: Workshop Summary}.
\newblock National Academies Press, 2009.

\bibitem{Hawe:2015ca}
Hawe P.
\newblock Lessons from complex interventions to improve health.
\newblock {\em Annual review of public health}, 36:307--323, 2015.

\bibitem{Datta:2013gf}
Datta J and Petticrew M.
\newblock Challenges to evaluating complex interventions: a content analysis of
  published papers.
\newblock {\em BMC Public Health}, 13(1):1, 2013.

\bibitem{Clark:2013hs}
Clark AM.
\newblock What are the components of complex interventions in healthcare?
  theorizing approaches to parts, powers and the whole intervention.
\newblock {\em Social Science \& Medicine}, 93:185--193, 2013.

\bibitem{Shadish:2002uv}
Shadish WR, Cook TD, and Campbell DT.
\newblock {\em Experimental and quasi-experimental designs for generalized
  causal inference.}
\newblock Houghton, Mifflin and Company, 2002.

\bibitem{Taljaard:2014kc}
Taljaard M, McKenzie JE, Ramsay CR, and Grimshaw JM.
\newblock The use of segmented regression in analysing interrupted time series
  studies: an example in pre-hospital ambulance care.
\newblock {\em Implementation Science}, 9(1):1, 2014.

\bibitem{Penfold:2013bc}
Penfold RB and Zhang F.
\newblock Use of interrupted time series analysis in evaluating health care
  quality improvements.
\newblock {\em Academic Pediatric Association}, 13(6):S38--S44, 2013.

\bibitem{ClinicalNurseLeade:2015tr}
Bender M, Murphy E, Thomas T, Kaminski J, and Smith B.
\newblock Clinical nurse leader integration into care delivery microsystems:
  Quality and safety outcomes at the unit and organization level.
\newblock In {\em Poster presented at Academy Health Annual Research Meeting,
  Minneapolis, MN}, 2015.

\bibitem{Bender:2012ch}
Bender M, Connelly CD, Glaser D, and Brown C.
\newblock Clinical nurse leader impact on microsystem care quality.
\newblock {\em Nursing Research}, 61(5):326--332, 2012.

\bibitem{Kavaanagh:2012dt}
Kavanagh KT, Cimiotti JP, Abusalem S, and Coty M.
\newblock Moving healthcare quality forward with nursing-sensitive value-based
  purchasing.
\newblock {\em Journal of Nursing Scholarship}, 44(4):385--395, 2012.

\bibitem{Bender:2014}
Bender M.
\newblock The current evidence base for the clinical nurse leader: a narrative
  review of the literature.
\newblock {\em Journal of Professional Nursing}, 30(2):110--123, 2014.

\bibitem{Rickles:2009}
Rickles D.
\newblock Causality in complex interventions.
\newblock {\em Medicine, Health Care and Philosophy}, 12(1):77--90, 2009.

\bibitem{Petticrew:2013}
Petticrew M, Rehfuess E, Noyes J, Higgins J, Mayhew A, Pantoja T, Shemilt I,
  and Sowden A.
\newblock Synthesizing evidence on complex interventions: how meta-analytical,
  qualitative, and mixed-method approaches can contribute.
\newblock {\em Journal of clinical epidemiology}, 66(11):1230--1243, 2013.

\bibitem{Kontopantelis:2015cn}
Kontopantelis E, Doran T, Springate DA, Buchan I, and Reeves D.
\newblock Regression based quasi-experimental approach when randomisation is
  not an option: interrupted time series analysis.
\newblock {\em British medical journal}, 350:h2750, 2015.

\bibitem{EPOC:2001tx}
Effective Practice and Organisation of~Care~(EPOC).
\newblock Interrupted time series (its) analyses.
\newblock {\em EPOC Methods PaperIncluding Interrupted Time Series (ITS)
  Designs in a EPOC Review}, 2001.

\bibitem{EPOC:2015ww}
Effective Practice and Organisation of~Care~(EPOC).
\newblock Interrupted time series (its) analyses.
\newblock {\em EPOC Resources for review authors. Oslo: Norwegian Knowledge
  Centre for the Health Services}, 2015.

\bibitem{Ramsay2003}
Ramsay CR, Matowe L, Grilli R, Grimshaw JM, and Thomas RE.
\newblock Interrupted time series designs in health technology assessment:
  lessons from two systematic reviews of behaviour change strategies.
\newblock {\em Int J Technol Assess Health Care}, 19(4):613--623, 2003.

\bibitem{quandt1958}
Quandt RE.
\newblock The estimation of the parameters of a linear regression system
  obeying two separate regimes.
\newblock {\em Journal of the american statistical association},
  53(284):873--880, 1958.

\bibitem{Thistlethwaite:1960wy}
Thistlethwaite DL and Campbell DT.
\newblock Regression-discontinuity analysis: An alternative to the ex post
  facto experiment.
\newblock {\em Journal of Educational psychology}, 51(6):309, 1960.

\bibitem{Cabrieto:2016ba}
Cabrieto J, Tuerlinckx F, Kuppens P, Grassmann M, and Ceulemans E.
\newblock Detecting correlation changes in multivariate time series: a
  comparison of four non-parametric change point detection methods.
\newblock {\em Behavior Research Methods}, pages 1--18, 2016.

\bibitem{Shardell:2007}
Shardell M, Harris AD, El-Kamary SS, Furuno JP, Miller RR, and Perencevich EN.
\newblock Statistical analysis and application of quasi experiments to
  antimicrobial resistance intervention studies.
\newblock {\em Clinical Infectious Diseases}, 45(7):901--907, 2007.

\bibitem{SAS}
SAS~User Guide.
\newblock Sas/ets 9.2 user’s guide, 2008.

\bibitem{Parienti:2011}
Parienti JJ, Cattoir V, Thibon P, Lebouvier G, Verdon R, Daubin C, and et~al.
\newblock Hospital-wide modification of fluoroquinolone policy and
  meticillin-resistant staphylococcus aureus rates: a 10-year interrupted
  time-series analysis.
\newblock {\em Journal of Hospital Infection}, 78(2):118--122, 2011.

\bibitem{Cable:2001}
Cable G.
\newblock Enhancing causal interpretations of quality improvement
  interventions.
\newblock {\em Quality in Health Care}, 10(3):179--186, 2001.

\bibitem{Mahamat:2007}
Mahamat A, MacKenzie F, Brooker K, Monnet DL, Daures J, and Gould I.
\newblock Impact of infection control interventions and antibiotic use on
  hospital mrsa: a multivariate interrupted time-series analysis.
\newblock {\em International journal of antimicrobial agents}, 30(2):169--176,
  2007.

\bibitem{Aboagye-Sarfo}
Aboagye-Sarfo P, Mai Q, Sanfilippo FM, Preen DB, Stewart LM, and Fatovich DM.
\newblock A comparison of multivariate and univariate time series approaches to
  modelling and forecasting emergency department demand in western australia.
\newblock {\em Journal of biomedical informatics}, 57:62--73, 2015.

\bibitem{Linden:2015vo}
Linden A.
\newblock Conducting interrupted time-series analysis for single-and
  multiple-group comparisons.
\newblock {\em Stata Journal}, 15(2):480--500, 2015.

\bibitem{Muggeo:2015}
Muggeo VMR.
\newblock Segmented relationships in regression models with
  breakpoints/changepoints estimation.
\newblock {\em CRAN-R 0.2}, 9, 2012.

\bibitem{Bhaskaran2013}
Bhaskaran K, Gasparrini A, Hajat S, Smeeth L, and Armstrong B.
\newblock Time series regression studies in environmental epidemiology.
\newblock {\em International journal of epidemiology}, page dyt092, 2013.

\bibitem{GareyLai}
Garey KW, Lai D, Dao-Tran TK, Gentry LO, Hwang LY, and Davis BR.
\newblock Interrupted time series analysis of vancomycin compared to cefuroxime
  for surgical prophylaxis in patients undergoing cardiac surgery.
\newblock {\em Antimicrobial agents and chemotherapy}, 52(2):446--451, 2008.

\bibitem{AnsariGray}
Ansari F, Gray K, Nathwani D, Phillips G, Ogston S, Ramsay C, and Davey P.
\newblock Outcomes of an intervention to improve hospital antibiotic
  prescribing: interrupted time series with segmented regression analysis.
\newblock {\em Journal of Antimicrobial Chemotherapy}, 52(5):842--848, 2003.

\bibitem{davis2006}
Davis RA, Lee TCM, and Rodriguez-Yam GA.
\newblock Structural break estimation for nonstationary time series models.
\newblock {\em Journal of the American Statistical Association},
  101(473):223--239, 2006.

\bibitem{Ombao:2015}
Kirch C, Muhsal B, and Ombao H.
\newblock Detection of changes in multivariate time series with application to
  eeg data.
\newblock {\em Journal of the American Statistical Association},
  110(511):1197--1216, 2015.

\bibitem{Shumway:2011}
Shumway RH and Stoffer DS.
\newblock {\em Time series analysis and its applications: with R examples}.
\newblock Springer Science \& Business Media, 2011.

\bibitem{granger2014}
Granger CWJ and Newbold P.
\newblock {\em Forecasting economic time series}.
\newblock Academic Press, 2014.

\bibitem{bollerslev1988}
Bollerslev T.
\newblock On the correlation structure for the generalized autoregressive
  conditional heteroskedastic process.
\newblock {\em Journal of Time Series Analysis}, 9(2):121--131, 1988.

\bibitem{bernal2016}
Bernal JL, Cummins S, and Gasparrini A.
\newblock Interrupted time series regression for the evaluation of public
  health interventions: a tutorial.
\newblock {\em International journal of epidemiology}, page dyw098, 2016.

\end{thebibliography}

\section*{Acknowledgments }

This study was funded in part by the Commission on Nurse Certification, and based upon work supported by the Eugene Cota-Robles Fellowship at the University of California, Irvine, the NSF Graduate Research
Fellowship under Grant No. DGE-1321846, and by the NSF MMS 1461534 and NSF DMS 1509023 grants. Any opinion, findings, and conclusions or recommendations expressed in 
this material are those of the authors(s) and do not necessarily reflect 
the views of the National Science Foundation.

\pagebreak

\appendix
\section{Appendices}

\subsection{A Note on the Time Series Length for Robust-ITS}
The number of time points required pre- and post-change point (or pre- and post-intervention) depend on many factors. Previously in the ITS literature, it has been suggested that a minimum of three time points is needed in both phases to adequately estimate the outcome means \cite{Ramsay2003, EPOC:2015ww}.

Estimating the intercept and slope of a straight line via regression requires at least three data points, to have sufficient degrees of freedom to estimate the variance. 
The constraint of three data points therefore makes the assumption that only an intercept and slope need to be estimated; not true here, since we also wish to model the correlation structure. Another data point is needed for each additional parameter estimated.
Ignoring the change point, since we are estimating the intercept, slope, autocorrelation, and variance of each segment, a minimum of five time points in each phase is needed to be able to merely estimate the parameters.

Because we estimate the change point it is necessary to obtain five time points in each phase separate from the set of possible change points to adequately estimate the regression lines. That is a total of 10 (5 for the pre- and 5 for the post-change point phases) plus the length of the set of possible change points, is required. 

The number of parameters that must be estimated plus one, is a severe lower bound for the number of time points needed to make inference and should not be used as a rule of thumb.

The discussion of setting a practical lower bound for the time points needed in each phase stems from the desire to have enough power to make proper inference. However, power not only depends on the length of the time series in each phase, but additionally on the distribution of the data points pre- and post-change point, variability, effect strength, and confounding \cite{bernal2016}. Considering solely the length of the pre- and post-change point phases is not sufficient when calculating power, many other factors must be taken into account. 
Little development of power calculations in the ITS setting exist \cite{bhaskaran2013}. 

\subsection{Mean Parameter Estimates}
The ordinary least squares (OLS) estimates for the mean parameters in $\theta$ of Section 3.2.1 are:
\begin{gather} 
\hat{\beta_0} = \, \bar{Y}_{1:(\hat{\tau} -1)} - \hat{\beta_1} \frac{\hat{\tau}}{2}, \label{MLE0} \\ 
\hat{\beta_1} =  \, \frac{\sum\limits_{t = 1}^{\hat{\tau} -1 } (t - \frac{\hat{\tau}}{2})\,Y_t}{\sum\limits_{t = 1}^{\hat{\tau} -1 } (t - \frac{\hat{\tau}}{2}) \, t}, \label{MLE1}\\
\hat{\delta} = \, \bar{Y}_{\hat{\tau}:T} - (\hat{\beta_1} + \hat{\Delta}) \, \bar{t}_{\hat{\tau}:T} - \bar{Y}_{1:(\hat{\tau} -1)} - \hat{\beta_1} \frac{\hat{\tau}}{2}, \label{MLE2} \\
\hat{\Delta} =  \, \frac{\sum\limits_{t = \hat{\tau}}^{T} (t - \bar{t}_{\hat{\tau}:T})\,Y_t}{\sum\limits_{t = \hat{\tau}}^{T} (t - \bar{t}_{\hat{\tau}:T})\,t}  \, - \,  \frac{\sum\limits_{t = 1}^{\hat{\tau} -1 } (t - \frac{\hat{\tau}}{2})\,Y_t}{\sum\limits_{t = 1}^{\hat{\tau} -1 } (t - \frac{\hat{\tau}}{2}) \,t}, \label{MLE3}
\end{gather}
where $ \bar{Y}_{a:b} =   \frac{\sum\limits_{t = a}^{b} Y_t }{ b-(a-1) } $ and 
% \,\,\,\,\,\,\,\, \text{and} \,\,\,\,\,\,\,\,   
$\bar{t}_{\hat{\tau}:T}  =  \frac{\sum\limits_{t = \hat{\tau}}^{T} t }{T - (\hat{\tau} -1) }. $ 
The estimates of $\beta_0$ and $\beta_1$ are the same as the OLS estimates obtained by fitting a linear model to the pre-change point phase alone. The estimates of $\delta$ and $\Delta$ may be obtained from fitting a linear model to the post-change point phase and subtracting the OLS estimates of the first phase from the OLS estimates (of the intercept and slope) of the second phase.
The estimates for $\sigma_1^2$ and $\sigma_2^2$ depend on the stochastic process, and are given in the following section for an AR(1) process. 

\subsection{AR(1) Parameter Estimates}
In the AR(1) setting with a change point at $\hat{\tau}$ the autocorrelation and variance can be estimated by maximizing the conditional likelihood. The estimates are functions of the residuals $R_t$ and the residuals of the residuals $W_t$
\begin{equation}
\hat{\phi} = \left\{ 
\begin{array}{lr}
\hat{\phi_1} = \frac{ \sum\limits_{t=2}^{\hat{\tau} -1} (R_t - \bar{R}_{2:(\hat{\tau} -1 )}) (R_{t-1} - \bar{R}_{1:(\hat{\tau} -2) }) }{ \sum\limits_{t=2}^{ \hat{\tau} -1} (R_t- \bar{R}_{2:( \hat{\tau} -1 )})^2} & \\
\\
\hat{\phi_2} = \frac{ \sum\limits_{t= \hat{\tau} + 1}^{T} (R_t - \bar{R}_{(\hat{\tau} +1):T}) (R_{t-1} - \bar{R}_{\hat{\tau}:(T-1)}) }{ \sum\limits_{t=\hat{\tau} + 1}^{T} (R_t - \bar{R}_{(\hat{\tau} +1 ):T})^2} &  \, \, \,   
\end{array}
\right.
\end{equation} 

%\vspace{.5mm}
\begin{equation} 
\hat{\sigma^2} = \left\{  
\begin{array}{lr}
\hat{\sigma_1^2} = \frac{1}{\hat{\tau} - 1} \sum\limits_{t=2}^{\hat{\tau}} \big[ (W_t - \overline{W}_{2:(\hat{\tau} -1 )}) - \hat{\phi_1} (W_{t -1} - \overline{W}_{1:(\hat{\tau} -2) })\big]^2 &   \\
\\
\hat{\sigma_2^2} = \frac{1}{T -  \hat{\tau} - 1} \sum\limits_{t= \hat{\tau} + 2}^{T} \big[ (W_t - \overline{W}_{(\hat{\tau} +1):T}) - \hat{\phi_2} (W_{t -1} - \overline{W}_{(\hat{\tau} +1):T}) \big]^2,  & 
\end{array}
\right.
\end{equation} 

\vspace{.5mm} \noindent where  $\bar{R}_{a:b}$ and $\overline{W}_{a:b}$ are the means of the residuals and of the residuals of the residuals, respectively, for time points $a$ through $b,$ and
$$W_t = \left\{
\begin{array}{lrr} 
R_t - \hat{\phi_1} \, R_{t-1},  &  & \,\, \, \,  \, \, \,\, \, \,  \, \, \,  \, \, \, \, \, \, \, \, \,  \, \, \, 1 < t < \hat{\tau}\\
R_t - \hat{\phi_2}  \, R_{t-1},  &   &  \hat{\tau} < t \leq  T. 
\end{array}
\right.
$$

\subsection{Nested F-test for the Equality of Autocorrelation for an AR(1)}
To determine whether the stochastic process differs as a result of the change point, we test the hypothesis that $\nu \equiv \phi_2 - \phi_1$ equals zero. 
If $\nu = 0$, there is one overarching AR(1) process for all time points, and equation \ref{AR1} reduces to
\vspace{-0.1in}
\begin{equation} \label{ARR}
\centering
R_t = \phi_1 \, R_{t-1} + e_t,  \,\,\,\,\,\,\, 1 < t \leq T,
\end{equation}
otherwise equation \ref{AR1} holds. 
We are comparing nested models where equation \ref{AR1} is the full model and 
equation \ref{ARR} is the reduced model, so an F-test is appropriate. 
The degrees of freedom corresponding to the reduced model is
$(T-1) -1,$ to account for the lag of the AR process and the parameter in the model, $\phi_1.$  Similarly, since the full model corresponds to two separately fit AR(1) processes and two parameters, the degrees of freedom is $(T-2) - 2.$
Denote the residual sum of squares for the reduced and full models, respectively, as
\vspace{-0.1in}
\begin{align*}
RSS_R = & \sum\limits_{t = 2}^T (R_t - \phi_1R_{t-1} )^2  \\
RSS_F = & \sum\limits_{t = 2}^{\hat{\tau} -1} (R_t - \phi_1 R_{t-1} )^2 + \sum\limits_{t = (\hat{\tau} +1)}^T (R_t - \phi_2 R_{t-1} )^2.
\end{align*}
Then the F-statistics is
$$ F = \frac{\frac{RSS_R - RSS_F}{([T-1] -1) - ([T-2] -2)}}{\frac{RSS_F}{(T-1) -1}} = \frac{(RSS_R - RSS_F)/{2}}{(RSS_F)/(T-2)},
$$
and under the null hypothesis ($\nu = 0$) is distributed $F_{2, (T-2)}$.

\subsection{Segmented Regression Models}
In the healthcare intervention literature there are two main types of segmented regression models utilized to model the trends. The first is parametrized in the same manner as \ref{EQ3}, with $\tau$ set to the time of intervention --- an assumed instantaneous effect --- that is, the mean is parametrized as:
\begin{equation} \label{EQ4}
\mu_{t}^1 = \left\{
\begin{array}{lr}
\beta_{0} + \beta_{1} \, t,  &  \,  t <  t^* \\
( \beta_{0} + \delta) \, + (\beta_{1} + \Delta) t,  &  t \geq t^* 
\end{array}
\right.
\end{equation}
where $t^*$ denotes the intervention time. For the data described in the Introduction, $t^* = 31.$
The second segmented regression model is
\begin{equation} \label{EQ5}
\mu_{t}^2 = \left\{
\begin{array}{lr}
\beta_{0} + \beta_{1} \, t,  &  \,  t <  t^* \\
( \beta_{0} + \psi) \, + \beta_{1} t + \Psi ( t - t^* +1 ),  &  t \geq t^* ,
\end{array}
\right.
\end{equation}
in which the time after intervention implementation is multyplying $\Phi$; as opposed to simply time, as in equation \ref{EQ4}.

Note, the trends prior to the intervention introduction are the exact same for both equations \ref{EQ4} and \ref{EQ5}. Post the intervention time, the intercept increase is denoted by $\delta$ in equation \ref{EQ4} and by $\psi - (t^* -1)\Psi$ in equation \ref{EQ5}, implying  $\delta = \psi - (t^* -1)\Psi$. The change in slopes is denoted by $\Delta$ and $\Psi$ in equation \ref{EQ4} and \ref{EQ5} respectively, and so $\Delta = \Psi.$ Although the parametrization is different, the estimates of the intercepts, slopes, and any function of the slopes and intercepts (as is the level change) are the same. Thus the models are equivalent. 

\subsection{Supporting Figures}

\vspace{-.2cm}
\begin{figure}[H] \centering
	\includegraphics[width=16.5cm,height=10cm]{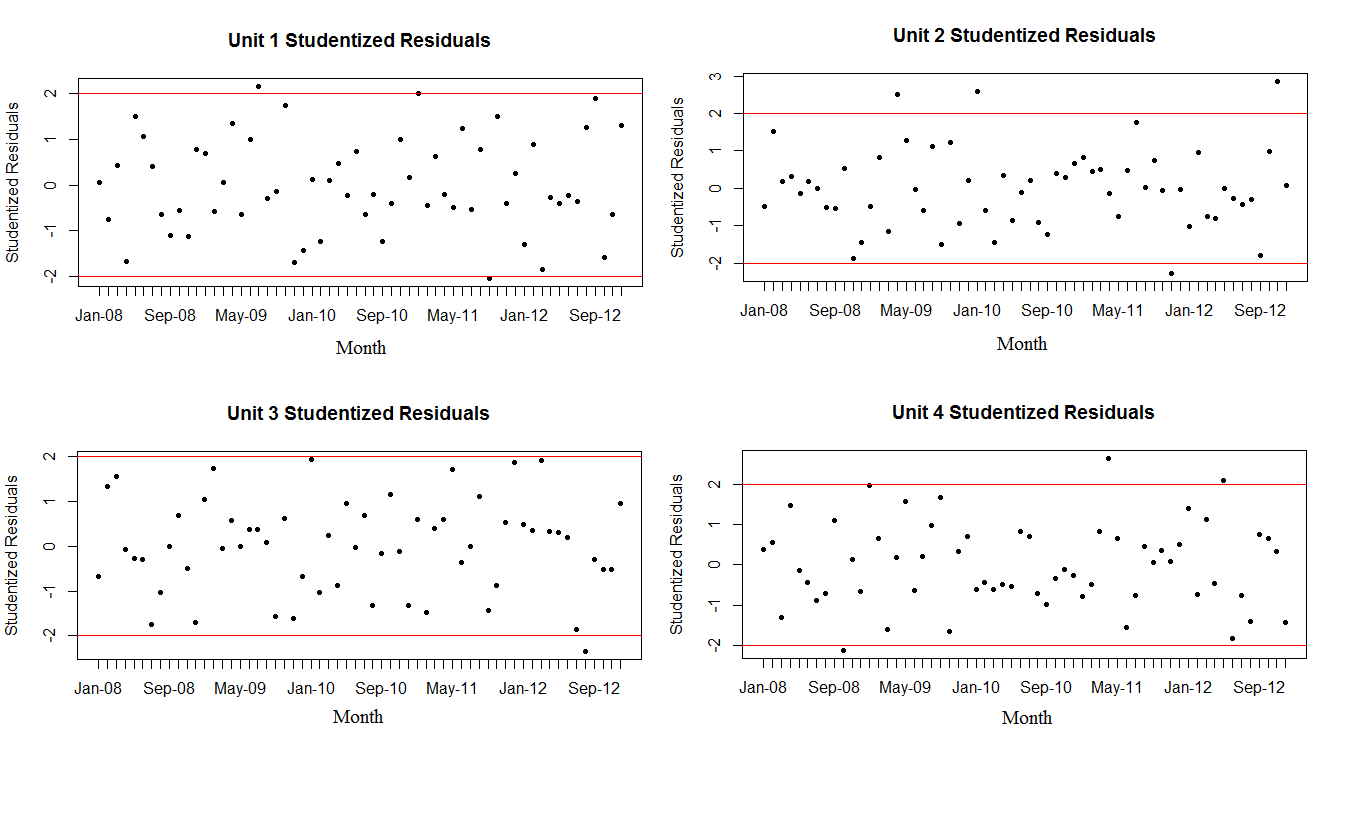} 
	\caption{Plots the studentized residuals of the Robust-ITS estimated patient satisfaction means for each unit. The studentized residuals do not exhibit any clear patterns, and seem to be closely centered around zero, indicating appropriate fits. The rule of thumb, ±2, is provided in each plot. }
	\label{Resplots}
\end{figure}

\begin{figure}[H] \centering
	\includegraphics[width=16.5cm,height=10cm]{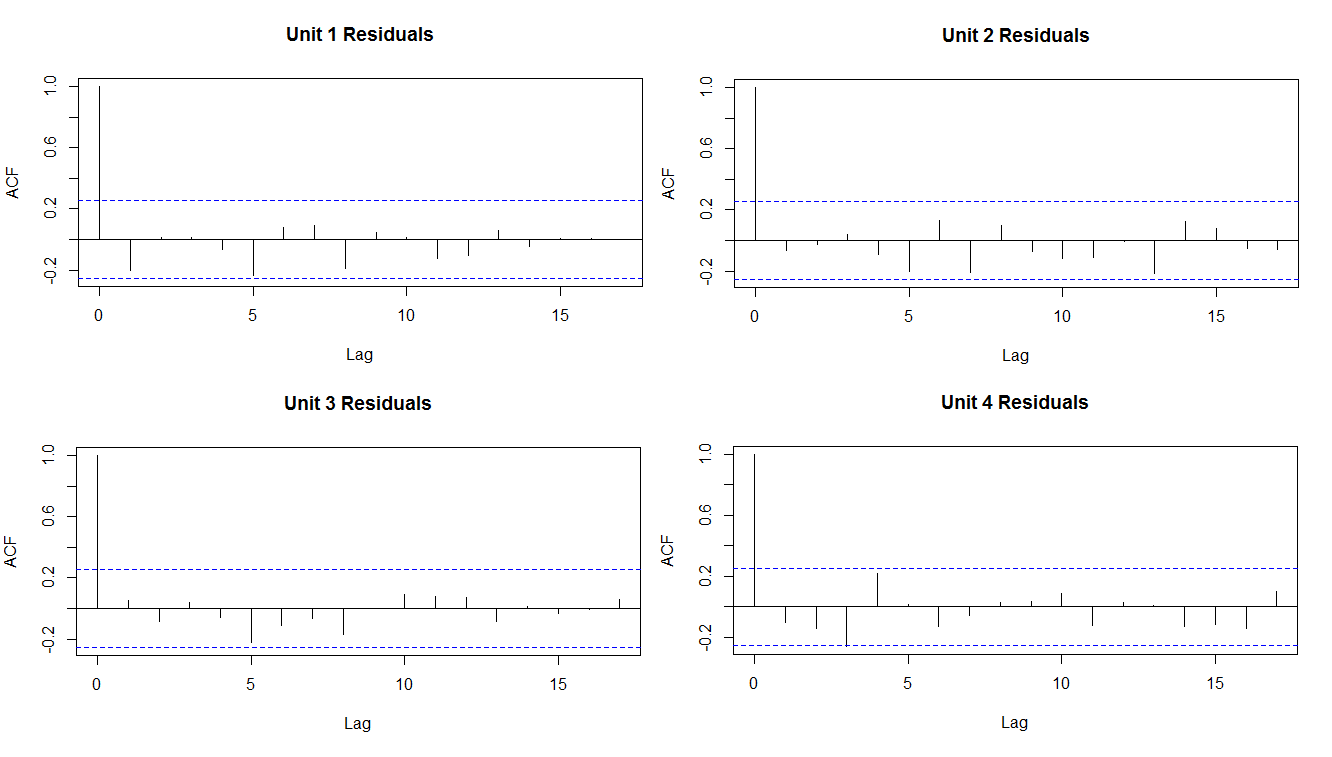} 
	\caption{Provides ACF, autocorrelation function, plots of the Robust-ITS estimated patient satisfaction means for each unit. The ACF plots suggest that the residuals behave as white noise, since the autocorrelation at lags greater than zero are small and seem to get closer to zero as the lag increases. }
	\label{ACFplots}
\end{figure}

%%%%%%%%%%%% Figure of fitted lines for Unit 1.

\end{document}